# Secure Handshake Mechanism for Autonomous Flying agents using Robust Cryptosystem

N Chandra Kanth[1*], Hemanth Rao K N[2**], Anjan K Koundinya[3]

[1, 2, 3] Department of Computer Science and Engineering, R.V College of Engineering, Bengaluru, India
[*]nckanth090@gmail.com
[**]h.rao46@gmail.com

**Abstract:** The autonomous flying agents in a Network centric environment and brings out various security threats and various techniques of Cryptography. Primary Focus is on study and implementation of how cryptographic algorithms can be effectively be used in a warfare scenario. The data security is utmost key factor for protection of data in such environments. The paper proposes mechanisms secured data transmission from command centre (which can be the sending flying agent) to shooter target. The command centre and shooter target has a unique set of encryption and decryption key which are created randomly by calibrating the security level at run time. At the beginning, the encryption key used for encrypting data is received from shooter target when the communication is authenticated through UDP sockets. The encrypted data is sent to the shooter target with the signed signature and command centre's encryption key. The encrypted data and signature is then decrypted and verified respectively at the shooter target. The time analysis is performed and observed inputs are provided to the command centre.

## 1. Introduction

In the advent of fast and mathematically complex cryptography mechanisms for reliable and secure information transfer among various devices ranging from small sensor to large commercial servers, there is strong emphasis on the encryption process adopted in functionalities of critical systems. Such critical systems are the core of Network centric warfare which has the technical superiority over enemy system of systems.

Netcentric warfare or also called as Network-Centric Warfare (NCW) is the dominant concept or hypothesis which is the driving force in the current makeover of military and military based affairs. These days there are a plethora of ways in which the militaries or the forces could be reorganized to make use of communication and networks more proficiently, and also in the scenarios where the networks themselves are made into structures or is made as a structured environment to work with, there is briskly emerging interest in the field of modelling and investigation so as a means for classification in the possibilities of warfare, and is therefore corresponding to the interest in metrics for NCW [1]. There are many reasons to consider this approach and proves that this is a clear cut solution, one of them being that not all potentials will lead to a full network-centricity, therefore it is essential for user's to be able to recognize and judge the net-centricity of the scenario or the environment and also its absence so that a better and effective military force is developed to associate its existence with improvements in military efficiency and effectiveness

Network-Centric Operations (NCO) on a warfare scenario is a crucial and important component of the Department of Defence. This is mainly for planning and transformation of the military operations [6]. Computer equipment and networked communications technology is the crucial tech on which NCO is dependent on to provide awareness in battle space [6]. Theoreticians argue that a communal cognizance increases the cooperation between command and control, stemming in excellent decision-making, and the skill to be able to coordinate complex and tough military operations over extended distances possible for an overwhelming war-fighting lead. Several DoD key programs and plans are underway for implementation across all services [5].

Protection of data is very important as any type of intervention in data can jeopardize a mission and may also lead to life threatening situations. Cryptography is the science introduced for protecting these data. It protects data from identity interception, repudiation, replay attacks and more. Cryptography is one of the main foundations of network-centric operations too.

In this paper, focus is given to the implementation of a secure handshake between computers in a network for data transmission using cryptographic algorithms. Also to perform the analysis on encryption time based on the data provided for transmission across a network, to detect and overcome possible chances of data tampering that can occur in a network by adversaries.

## 2. Related Work

Internet provides essential communication between billions of people and is being increasingly used as a tool for commerce. Security becomes a tremendously important issue to deal with. There are many aspects to security and many applications, ranging from secure commerce and payments to private communications and protecting passwords. One essential aspect for secure communications is cryptography.

In [11] it is discussed that low power synthesis is an advantage of AES but More reversible gates are required than the traditional CMOS designs which is a limitation. In [12] Avalanche effect in AES algorithm is discussed in detail where it is found out that high avalanche effect results in improved security level which is an advantage but there is a drawback that it cannot carryout experiments on image. In [5] GPU; style of stream programming is discussed where it is found out that there is improved encryption computing efficiency of the algorithm using this stream programming, where as there is a limitation in which The speed of the CPU cipher algorithms is determined by the standards of openssl command. In [13] analysis of AES and RSA is performed where it is found out that AES encrypts the data blocks faster than RSA Reduced encryption with large keys, resulting is less compromising of the keys. Limited damage if the key is compromised. The following is found out to be a limitation. (1) Distribution of the secret key is an issue with AES. (2) High computational overhead is involved with RSA.

## 3. Mathematical Foundation

### 3.1. Finite Field Cryptography

Mathematical Foundations used in Cryptography: -

Field: - is a set of numbers in which we can add, subtract, multiply and divide.

In Cryptography we almost always need finite fields.

Theorem: Finite Field (F.F) or Galois Field (GF) only exists if they have pm elements.
p = prime.    m = positive integer.

For example,
There is a finite field with 11 elements.   **GF (11)**

There is a finite field with 256 elements.   **GF (2^8)**

There is no finite field with 12 elements. **GF (2^2.3)**

Types of Finite Field are (see Fig.1):

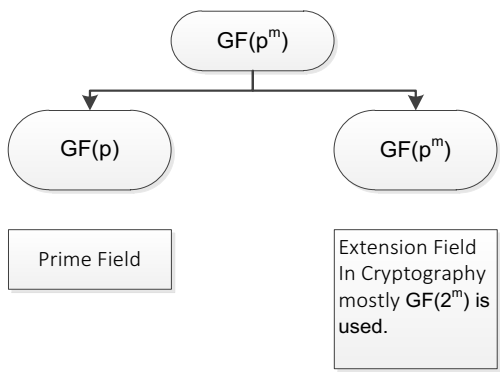

*Fig.1.* Classification of Finite FieldPrime Field

The elements of a prime field G(p) are integers {0, 1, p-1} and does following operations {+, -, x, ()$^{-1}$}

$ADD: a + b = c \bmod p, SUBTRACT\ a - b = d \bmod p$

$MULTIPLICATION: ab = e \bmod p^{-1}, aa^{-1} = 1 \bmod p$

Extended Euclidean Algorithm is used to compute the inverse of a number modulo n.

### 3.2. Homomorphic Encryption

The lineage of the word homomorphic is to the Greek language which translates as the same form or the same shape, and the core concept of it is the transformation that has the same effect on two different sets of objects.

The principle of homomorphic encryption is a base on abstract algebra term homomorphism, which refers to mapping between two groups (G, *) and (H, *) such that:

$$(x * y) = x * y\ \forall\ x, y\ \in G\ and\ x\ \in H$$

## 4. Visualization and System Design

### 4.1. Requirements of Netcentric Operation Data Security

1. *Privacy:* is the technique of keeping sensitive information private and secret, so that only the intended recipient which the user acknowledges is able to understand the information.
2. *Validation*: is the process of delivering and providing the proof of identity of the sender to the recipient. This is so that the recipient can be guaranteed that the individual sending the information as a data is who and what he or she claims to be.
3. *Integrity*: this is the technique to make sure that information is not corrupted or rigged during its transit or the storage of that information on a network. An unauthorized individual should not be able to tamper and change the information during transit.
4. *Non-Repudiation*: is the technique to make sure that information or data is not being disclaimed and rejected which means, that the sender cannot and should not be denied as being the creator of the information or data once it is received by the recipient.

### 4.2. Architecture for an Adaptive Ad-Hoc Netcentric Environment

The architecture of our system as shown (see Fig.2) is deployed considering the Observe, Orient, Decide, Act (OODA) Loop. Where we majorly focus of the Inference, Detection and Act cycle of the Loop.

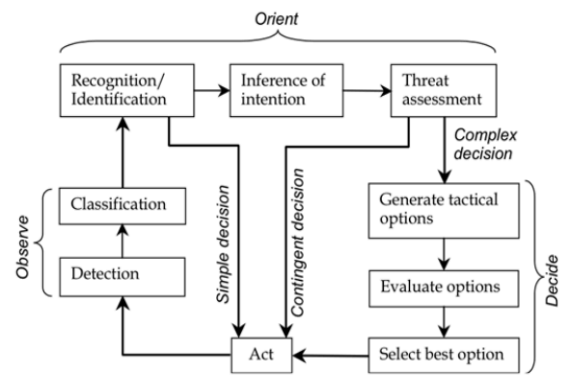

*Fig.2.* The OODA Loop Decomposed

Architecture of an adaptive ad-hoc network supporting network-centric operations describes the fundamental network elements and their interconnections. The architecture distinguishes two layers: infrastructure and logical. The logical layer describes mechanisms supporting the network adaptation functions necessary for efficient functioning of network elements. Network elements are defined in the infrastructure layer. For the assumed ad-hoc network architecture, the fundamental network elements are user terminals responsible for providing services to users and also performing the role of intermediary nodes in the data exchange process. The selected user terminals can also be a gateway to the external network (by launching additional functionalities). For our network concept we assumed a use of wireless LAN standard based on standard IEEE 802.11s for communication networks elements. The IEEE 802.11s standard - "ESS - Extended Service Set Mesh Networking" assumes the possibility of creating a self-configuring 802.11 wireless network operating in an ad-hoc mode, where each device can have several interfaces. One of the basic assumptions of the developed solution was to provide extensibility and flexibility, applying a variety of mechanisms to fulfil the same functions in different nodes of the network.

For an IEEE 802.11s standard the use of the underlying mechanisms for the physical layer is assumed, that is: the security of information and access to the transmission medium used in previous versions of IEEE 802.11. Additionally, the use of new additional solutions is assumed. The required functionality described by the IEEE 802.11s includes:

• *PHY* - Physical Layer, based on the IEEE 802.11a /b/g/n.

• *Medium Access Coordination* - mechanisms of medium access control based on standard solutions adopted for the network compatible with IEEE 802.11 with QoS extensions described in the IEEE 802.11e. For further use new features are expected such as the ability to dynamically change the operating channel.

• *Mesh Security* - authentication of users and nodes as well as cryptographic trunks protection, mainly based on the extension of IEEE 802.11i.

• *Mesh Configuration & Management* - mechanisms of automatic configuration of radio parameters (operating channel frequency selection, transmitting power, etc.), QoS management policies.

• *Discovery & Association* - mechanisms of detecting the presence of neighbouring nodes and a mesh network as well as identifying parameters of a mesh network, the procedures of connecting nodes to the network and the states of logical links with its neighbours.

• *Mesh Topology, Learning, Routing & Forwarding* - a group of mechanisms and protocols for the control and the creation of the current topology and data forwarding. For this purpose, the use of a Hybrid Wireless Mesh Protocol (HWMPA) for routing [14] is assumed, which is a combination of mandatory reactive protocol mechanisms ad-hoc Radio Metric Ad-hoc On-demand Distance Vector (RM-AODV) [16] and optional proactive mechanisms - Tree Based Routing (TBR) [14]. It is also assumed to use a proactive routing protocol Radio Metric Optimized Link State Routing (RM-OLSR).

• *Internetworking* - mechanisms of ensuring mesh network cooperation with external 802 series networks, including the wired networks.

• *Mesh Measurement* - mechanisms and procedures for monitoring a mesh network and radio environment, e.g. for setting routes in a mesh network.

Stand-alone system is a system which are in no way connected or has a receiving connection with another system or any networking devices, this can be a switch, or a router with an exclusion of a directly-attached printer, or to a Metropolitan Area Network (MAN), Wide Area Network(WAN) or a Local Area Network (LAN). These systems or computers should run a vendor-supported (i.e., currently being patched) operating system, such as a current version of Windows workstation or a mainstream server. It can also include server operating systems like Linux, UNIX, or Macintosh OS X. These stand-alone systems are not connected to the Internet or a LAN or a WAN. The prominence is on data security which is can be and has to be placed on the security layer of the user's computer or system and this also helps to control the access to the data on the system which the stand-alone systems are connected.

The following are the steps to secure confidential data on a stand-alone system:

• Make sure that the computer boots from the hard drive only, this can be done by configuring the BIOS settings in the computer boot menu.

• Continuing from the previous point do not permit the stand-alone system to be booted from a CD Diskette or a drive.

• Authenticate and Master Lock the BIOS menu, so that password changes can be made only on the consent of the super user.

• Manually secure the system or the stand-alone computer by a table lock and cable. Also, secure the system by setting up a user password or a BIOS password as described earlier.

• Securely store all the data on a desktop computer only.

• Network Interface Card also called the NIC, is the only way a system can be connected to the Internet. Therefore, remove the NIC card so that no Internetworking can occur.

Other ways to remotely controlling the access of data:

• Use the security features available in the stand-alone operating system to restrict access to most of the sensitive data and important data which reside on such systems. This features can include login only through a secure user-id and a password and NTFS agreements on Windows systems and ACLs in Linux and OS X [8].

• Strong passwords are a must. Passwords must contain different characters of varying cases like (A, s, D, f), must contain special characters like ($, #, @) and numbers in an equal mixture of characters.

• There are many Executive tools and several local security policies to make passwords more complex.

• Get the passwords verified by the Dean and the Head of the Department or Director, so that the audit can verify your passwords with L0PHTCRACK.

• Screen savers are way of securing the system if the user forgets to lock the system manually, the screen saver can be turned on after 5-6 minutes of inactivity. Now, since the screen saver activity will not be activated until the 5-6 minutes of inactivity, it is suggested that the system user must manually lock the screen (Windows = W Key + L) every time the user walks away from the system.
• Use a Microsoft SCM Standard for safely securing the password or picking the right password.
• Use an encryption mechanism to encrypt the entire secure data on the stand-alone system, this can include whole-disk encryption or a directory based encryption some possible examples include Windows Encrypting File System or Veracrypt [12].
• The temporary work files which the user works with, can be made to be saved and pointed to the encrypted sensitive data directory.

The customary classification of various security dangers or threats are identified into three main classes. These depend on various factors and also on the stand-alone system property which is being threatened which are privacy, veracity or accessibility.
• Privacy is also called as confidentiality and is sullied when unofficial people also called hackers learn the protected information, and in the case of this paper in the field of communication between two standby computers.
• Veracity is sullied when unofficial people modify sensitive information, this can be when this person changes the amount of the recipient on a check.
• Accessibility is sullied when the stand-alone system is prohibited from executing its intended purpose or function, as when a hacker brings the Web-site of an online store is brought down.

The protection of the systems all rely on the user's distinction between authorized and unauthorized principals. Differentiating between these principals usually comprises of a three-step procedure:
• Identification,
• Authentication,
• Authorization.

Authentication failure can mainly lead to a defilement in the procedure of veracity, accessibility and privacy. For example, if the true identity of the recipient of the user is not anticipated by the user, then there is no point in encrypting the system and software so that the recipient can access it through unfair means. So it is often more likely that a given task of safely protecting and securing a new stand-alone computing environment- is to look at authentication first as shown in (see Fig.3).

The two stand-alone systems will act like the senders and receivers as it system allows Duplex communication and these two systems act like the two stand-alone systems in a ubiquitous environment. Starting with the system which wants to communicate with the receiver, she asks for the Encryption key from the receiver this means the selection of the proper encryption protocol and asking for its encryption key, which will be its public key. The receiver sends the key back to sender and she encrypts the message and assigns her signature which also will be encrypted and sent to the receiver.

The receiver now receives the message which is encrypted. He then decrypts the message, using his private key to reveal the message.

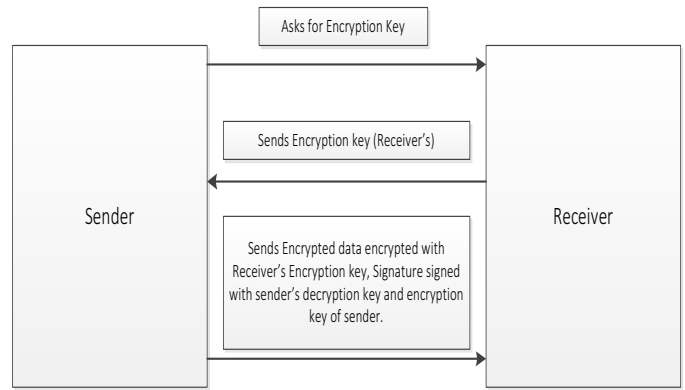

*Fig.3*: Data transfer between two entities

We call this the digital Signature Protocol

1. Alice signs the message encrypted using Bob's encryption key ($E_B$). By signing we mean decrypting the encrypted message here. Alice signs the message user her decryption key ($D_A$) which is private.

M- Original Message

$$M1 = D_A(E_B(M))$$

Alice sends M1 to Bob.

2. Bob verifies M1 using Alice's Encryption key (EA). (Note all encryptions are public) By verifying we mean encrypting the message here. After verifying decrypting, the message using his own private key. So,

Verifying,

$$E_A(D_A(E_B(M))) = E_B(M)$$

Now Decrypt,
$$D_B(E_B(M)) = M$$

Now, it's impossible for Eve to act as Alice unless he somehow gets Alice's decryption key. Hence by using this Alice and Bob can authenticate each other.

Unhindered exchange of information and communication across systems are the key elements for Network Centric Warfare and C4I. In evolution process of "need to know culture" to "need to share culture", high priority is given securing voice messages and sensitive data in a network during transmission and storage of data in databases. This data is subjected to consecutive changes to match the need of particular organization and to keep data safe from being exploited by external groups. Possession of data and keeping it safe and secure is a challenging task. Availability of the data without it being tampered at all times is crucial for success of every mission in armed forces (see Fig.4).

The following features encompasses this protection at upper most level:
• Securing the whole functional network depending on coherent security architecture of a wide spread system with secured zones and zone transition.

• Preservation of privacy, integrity and traceability of every sensitive data.
• Providing a provision for immediate support for all control and command information.
• In a heterogeneous system construction of homogenous and impermeable security layer at the upper layers of OSI model.
• Secure data links and voice, especially for messaging, IP VPN and radio. Interface all of this secure links on to the backbone of all the technologies.
• Hence, Cryptography plays an important role in taking care of these data security needed.

**Command Centre**
• There are 4 zonal command centre and 1 main command centre in India.
• Role: (Nodes: - Aerostats, Helicopter, Aircraft, AWACS, UAV)
• Command centre process the data received from several nodes
• Extract the information
• Creates the Battle field or Threat scenario
• Perform Decision making and disseminate commands to nodes.

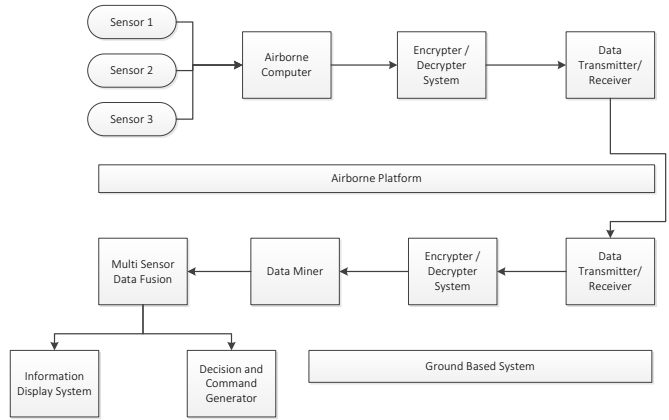

*Fig.4*: How a Data is transmitted

**Types of Data Transmitted**
• Geographical information: Satellite images from the military satellites.
• Airborne radar tracks
• Airborne radar images
• Voice based target information (Air, Land, and Sea)
• Sonar Information

The acceptance and incorporation of enhanced trending technologies, concepts, competencies and processes into preparation and operative activities will establish fore coming success. An organization must consist of a robust network which can support real time operation demands and should also provide a foundation of the investments with which that organization can grow and evolve as per needs of technology in current scenario. As agile, networked collaborative environment which is growing rapidly, these infrastructures are preferred over sequential, compartmentalized and point to point. Ubiquitous solutions are replacing point solutions. Interoperable and open systems are replacing proprietary solutions. As a result, a flexible, robust, resilient and secure network is desired.

Although there are many definitions for Network Centric Warfare (NCW), most of all approve that the fundamental principle is to hold an advantage in battle field by the use of network technology. ICRS defines Network Centric Operations (NCO) as the procedures 'on computer equipment and networked communications technology to provide a shared awareness of the battle space for Indian forces', this definition varies for different countries.

The case will be made that to increase the likelihood of success in operations and to enhance a military's war-fighting capability, a net centric approach to military operations is supreme.

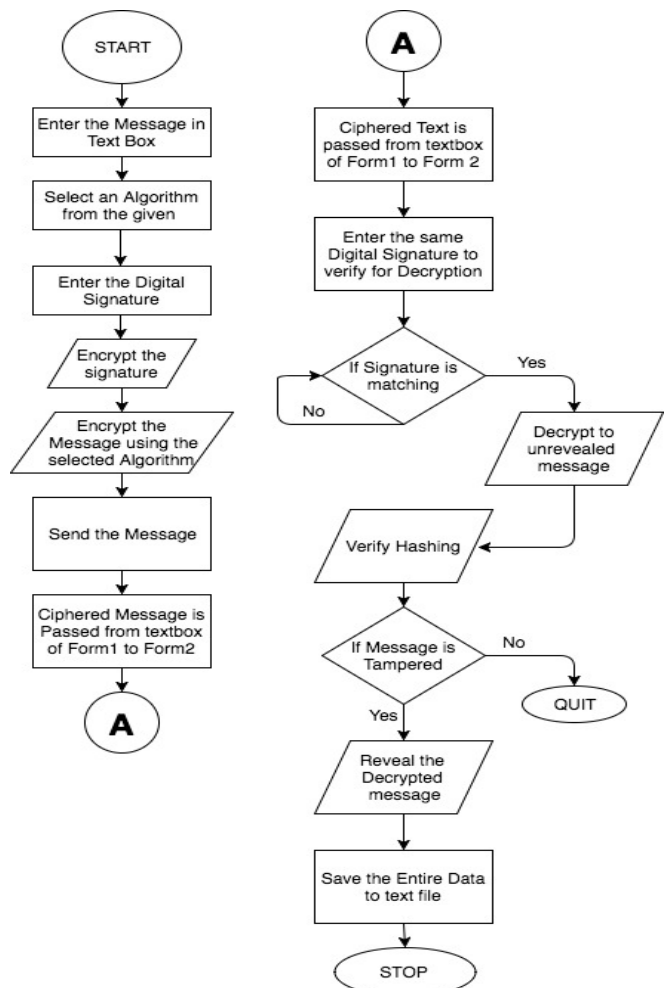

*Fig.5:* Flow chart for RSA, AES, DES and SHA-1 implementation

## 5. Experimental Analysis and Results

An easy way to comply with the requirements stated in the Author Guide [1] is to use this document as a template and simply type your text into it. PDF files are also accepted, so long as they follow the same style.

Experimental result of the Encryption algorithm which are AES, DES and RSA are shown in the bar graph below. These graphs show the contrast between the three algorithm using the same text for five investigations, these investigation is performed for a time which is taken into account for analysis and output byte for AES and DES is identical for different scopes of files. The time which was taken into account earlier is now checked and is seen that RSA algorithm takes a much longer time to produce the results when compared by the time taken by DES and AES algorithm. A large variation in the usage of memory for these various algorithms is also observed, but this variation does not increase as per the size of the files in algorithms.

Studying the results shown in (see Fig.6) which shows the time taken by these three algorithms i.e. AES, DES and RSA for encryption on various text sizes, it is found out that the RSA algorithm usually takes a longer processing time to compare than the time taken by DES and AES algorithm. DES and AES algorithm are seen to be showing a very minor variance in the time taken for encrypting the text. DES algorithm consumes least time for encryption.

Then, (see Fig. 7) which show how the memory is used up by DES, AES and RSA algorithm. It can be noticed that RSA algorithm uses more memory and is the peak for all sizes of document, while usage of memory is least

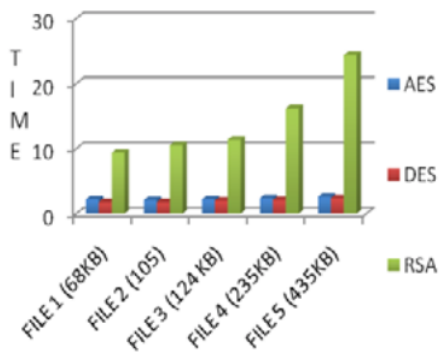

*Fig 6:* Comparison of Computing Time among AES, DES and RSA

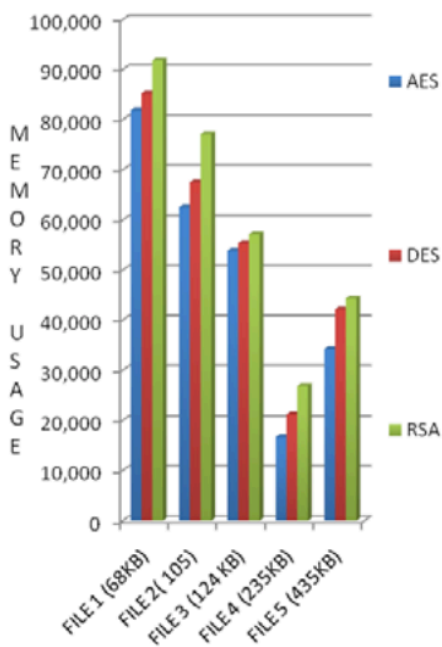

*Fig 7*: Comparison of memory usage by AES, DES and RSA

**Table.1** Encryption Time of the algorithm

| Test No | Amount of text to be Encrypted | Time taken by using AES in Encrypting data (Including time taken for generating random keys for each session) |
|---|---|---|
| 1. | 10 | 5.9837 seconds |
| 2. | 100 | 6.1670 seconds |
| 3. | 500 | 6.1170 seconds |
| 4. | 1000 | 6.4890 seconds |
| 5. | 1500 | 6.5170 seconds |
| 6. | 2000 | 6.5098 seconds |

After a clear study on security level, performance and efficiency of various algorithms and their effectiveness of use we havedeterminately found out RSA to be more secure to use for encryption of data. We have analyzed the time taken by RSA algorithm to encrypt data in our implementation of algorithm in Visual Studio, the order of time taken is as follows.

## 6. Conclusion

In a Network centric environment, the encryption algorithm used plays an important role. Our research work surveyed the existing encryption techniques like AES, DES and RSA algorithm. These encryption techniques are studied, analysed well and implemented to promote the performance of the encryption methods also to ensure the security. Based on the experimental result it was concluded that AES algorithm consumes least encryption, decryption time and buffer usage compared to DES algorithm, whereas RSA consume more encryption time and buffer usage is also very high. We also observed that decryption of AES algorithm is better than other algorithms. From the simulation result, we evaluated that AES algorithm is much better than DES and RSA algorithm.

It is concluded that, for reliable and secure data transfer among various avionics systems of autonomous Flying agents, it is necessary to encrypt/decrypt such data transfer using 32-bit or 64-bit AES encryption/decryption algorithm, which makes the Flying agents jam resistant to achieve the desired functionalities in tactical, operational and strategic domains of warfare.